\documentclass[epj]{svjour}
\usepackage{graphics}
\usepackage{amsmath}
\usepackage{amssymb}
\usepackage{epsfig}
\newcommand{\bel}[1]{\begin{equation}\label{#1}}
\newcommand{\bal}[1]{\begin{eqnarray}\label{#1}}

\newcommand{\expo}{\textrm{e}}
\newcommand{\imag}{\textrm{i}}

\begin{document}
\title{The partition function of an interacting many body system}
\subtitle{Beyond the perturbed static path approximation (PSPA)}
\author{Christian Rummel\inst{1} \and Joachim Ankerhold\inst{2}}
\institute{Physik-Department der Technischen Universit\"at M\"unchen, 
           James-Franck-Strasse, D-85747 Garching, Germany 
      \and Fakult\"at f\"ur Physik, Albert-Ludwigs-Universit\"at Freiburg, 
           Hermann-Herder-Strasse 3, D-79104 Freiburg, Germany}
\date{Received: date / Revised version: date}
%
\abstract{
Based on the path integral approach the partition function of
a many body system with separable two body interaction is calculated
in the sense of a semiclassical approximation. The commonly used Gaussian
 type of approximation, known as the  
perturbed static path approximation (PSPA), breaks down near a 
crossover temperature due to
instabilities of the classical mean field solution.
It is shown how the PSPA is systematically improved within the
crossover region by taking into account large non-Gaussian
fluctuations and an approximation applicable down to very low
temperatures is carried out.
 These findings are tested against exact results for the
archetypical cases of a particle moving 
in a one dimensional double well and the exactly solvable Lipkin-Meshkov-Glick
model. The extensions should have
applications in finite systems at low temperatures as in nuclear
physics and mesoscopic systems, e.g.\ for gap fluctuations in nanoscale
superconducting devices previously studied within a PSPA type of
approximation.
} 
\maketitle
\section{Introduction}
\label{intro}

Thermodynamic properties of interacting many body systems are of fundamental
interest for all kinds of  condensed matter. A
particular challenge has been the temperature range where strong
quantum effects render simple mean field theories
insufficient. 
For finite systems with separable
two-body interaction much can be gained by relying on approximate
techniques to evaluate the partition function. Typical examples include
 certain 
mesoscopic systems, e.g. ultrasmall 
superconducting metallic grains \cite{nanopart}
on which  substantial research has focused recently.

A very elegant approach to approximate the partition function is
provided by the path integral formalism.  
There, the usual trick \cite{negelej.orlandh,callawayj} is to turn the
two-body interaction via a Hubbard-Stratonovich transformation into
terms containing only one-body operators and an auxiliary field. While
the static part of this field determines the static mean field result, 
many body quantum fluctuations are encoded in its dynamical part. The
latter one  can
systematically be accounted for in the sense of a semiclassical
approximation. Accordingly, a static approximation has been developed 
originally in order to study finite size effects in small superconductors 
\cite{mub.scd.der:prb:72}. The static path approximation (SPA) has also 
been used in nuclear physics to calculate thermodynamic properties and 
level densities of hot nuclei \cite{aly.zij:prc:84}.
Later it has been extended by the SPA+RPA  
\cite{pug.bop.brr:ap:91}, the perturbed static path approximation (PSPA, 
the name which we take over in the following) \cite{ath.aly:npa:97}, 
or the correlated static path approximation (CSPA) 
\cite{ror.can:plb:97,ror.rip:npa:98}. These approaches account for
small Gaussian fluctuations around the classical mean field and are
thus applicable in a temperature range
where quantum effects are no longer negligible. This way, from the
approximate partition function a variety of thermodynamic quantities 
have been calculated in the literature: the free and 
internal energy, the specific heat and the level density of the system 
\cite{pug.bop.brr:ap:91,ath.aly:npa:97,agb.ana:plb:98,agb.sas.dej.shs:prc:98}. 
In addition thermal   
expection values of observables \cite{ath.aly:npa:97} and strength functions 
\cite{ath.aly:npa:97,ror.rip:npa:98} have been deduced.  While many of these 
calculations have been carried out in exactly solvable models 
\cite{pug.bop.brr:ap:91,ath.aly:npa:97,ror.can:plb:97,ror.rip:npa:98}
also  reasonable  agreement with experimental data was shown for the 
level density of finite nuclei \cite{agb.ana:plb:98,agb.sas.dej.shs:prc:98}.
Further, the formalism has been put forward in
\cite{agb.sas.dej.shs:prc:98} to take into account the coupling to  
the continuum and in \cite{ruc.hoh:pre:01} an extension 
of the PSPA to describe dissipative decay out of metastable
states has been developed in a self-consistent fashion. 
Very recently, the improvements of the SPA have been applied to small
strongly correlated condensed matter systems. 
For example,  using the CSPA odd-even effects in small superfluid
systems  \cite{ror.can.rip:ap:99} and 
motivated by new experiments on ultrasmall 
superconducting metallic grains \cite{nanopart}
 gap fluctuations and pairing effects in finite size superconductors 
have been examined \cite{rossignoli}. 

As noted above, these conventional improvements of the SPA take into
account quantum 
effects on the RPA level only, i.e.\ Gaussian fluctuations,  and thus,
are plagued by the problem that  
they break down at a certain temperature $T_0$.
Physically, at this temperature the classical mean field solution
becomes unstable in functional space and large fluctuations render the
Gaussian approximation insufficient. This has some features in common  
\cite{hoh.kid:ijmpe:98,ruc.hoh:pre:01} with the change of stability at a 
 crossover temperature occuring in the description of dissipative
decay e.g.\ in Josephson junctions \cite{devoret} but is different in detail.
 By thoroughly analysing the
appearence of instabilities for the many body problem at lower
temperatures, we go beyond  the PSPA  to achieve a smooth behavior
around $T_0$. For temperatures sufficiently below $T_0$ the partition
function is dominated by field configurations away from the classical
mean field around which  a Gaussian approximation is again valid. In
this paper we lay out the general theory and discuss its validity in
comparison to exactly solvable models before it can be applied to
realistic systems.

We start with a brief outline of the main results of the
conventional PSPA and introduce the notation used in the sequel
(Sec.~\ref{partfunc}). The extension of the PSPA around the critical
temperature 
$T_0$ is developed in Sec.~\ref{extPSPA} which in Sec.~\ref{lowtemp}
allows for a treatment of the 
temperature range far below $T_0$. In Sec.~\ref{1dim} our results
are first applied to a one-dimensional double well, 
before in Sec.~\ref{manybody} we 
turn to the thermodynamic properties of the Lipkin-Meshkov-Glick model
\cite{lih.men.gla:np:65}.

\section[Partition function]
{Partition function of an interacting many body system}
\label{partfunc}

In this paper we want to approximate the partition function of a system 
described by a Hamiltonian of the following structure
\bel{twobodham}
\hat{\cal H} = \hat{H} + \frac{k}{2} \,\hat{F}\hat{F} \ .
\end{equation}
Here $\hat{H}$ and $\hat{F}$ are hermitian one body operators and
the product $\hat{F}\hat{F}$ mimics an effective separable two
body interaction. In the sequel, for the general analysis we always assume
$[\hat{H},\hat{F}]\neq 0$, the simplification to the case of
commuting operators is straightforward.
Further, the coupling constant $k = -|k|$ is taken to be negative (attractive 
interaction) \cite{bohra.mottelsonb.2} according e.g.\ to isoscalar
modes. The effect of repulsive  
interaction has been studied in \cite{can.ror:prc:97}. 
The ansatz (\ref{twobodham}) 
defines a minimal microscopic model for a system with one collective
degree of freedom \cite{bohra.mottelsonb.2}.
An extension of the methods proposed 
in this paper to systems with more (independent) collective degrees of 
freedom is then feasible, though it may be tedious in detail.

The partition function of the grand canonical ensemble reads
\bel{Z}
\mathcal{Z}(\beta) = \textrm{Tr} \,\exp \left( -\beta (\hat{\cal H} -
\mu \hat{A}) \right) = \textrm{Tr} \ \hat{U} \ , 
\end{equation}
where $\beta = 1/T$ is the inverse temperature (in units with 
$k_{\textrm{B}} \equiv 1$) and the chemical potential $\mu$
keeps the particle number 
$\langle\hat{A}\rangle$ fixed on average. In principle, one should
work with truly fixed particle numbers, i.e. within the 
canonical ensemble. However, as we
are mainly interested in the dependence of system properties on
excitation energy or temperature, it is more convenient to exploit
(\ref{Z}). 
A very elegant way to do this is to represent 
${\cal Z}(\beta)$ as a functional integral in imaginary time
\cite{negelej.orlandh}, which also allows to systematically include
fluctuations around the mean field. Given the form of the Hamiltonian 
in (\ref{twobodham}) the mean field approximation starts with a 
Hubbard-Stratonovich transformation \cite{hubbstra} of
the imaginary time path integral corresponding to (\ref{Z}). Accordingly,  
the product $\hat{F} \hat{F}$ is split by introducing an auxiliary path 
$q(\tau)$ as a collective variable. Since this procedure is well-known 
(see e.g. \cite{ath.aly:npa:97,negelej.orlandh,callawayj}) we simply state 
here the basic results which will then serve as the starting point for our 
analysis 
(\footnote{We note that our notation has been adapted to the one used in 
\cite{hoh:pr:97,ruc.hoh:pre:01} and relates to that of \cite{ath.aly:npa:97} 
in the following way: 
$\hat{H}_{0} \leftrightarrow K$, $\hat{F} \leftrightarrow V$,
$k < 0 \leftrightarrow -\chi < 0$, $q \leftrightarrow \chi\sigma$,
$z \leftrightarrow \zeta_{0}$, ${\cal C} \leftrightarrow \zeta_{0}'$,
$\hbar \leftrightarrow 1$.}). 

After introducing the Fourier expansion 
\bel{fluctcoord}
q(\tau) = q_{0} + \sum_{r \ne 0} q_{r} \,\exp (\imag\nu_{r}\tau) \ ,
\qquad q_{-r} = q_{r}^{*} \ ,
\end{equation}
with the Matsubara frequencies
\bel{Matsubara} \nu_{r} = \frac{2\pi}{\hbar\beta} \,r \equiv
\frac{2\pi}{\hbar} \,r T \ \ , \qquad r = \pm 1,~\pm 2,~\pm 3 ~\dots,
\end{equation}
the partition function may be written in a form containing a static,
i.e.\ $q_0$ dependent, part and a dynamical factor, namely,
\bel{Z-athaly}
\mathcal{Z}(\beta) = \sqrt{\frac{\beta}{2\pi |k|}} \int_{-\infty}^{+\infty} 
\!\! dq_{0} \ \exp [-\beta \mathcal{F}^{\textrm{SPA}}(\beta, q_{0})]
\ C(\beta, q_{0}). 
\end{equation}
Here
\bel{FSPA}
\mathcal{F}^{\textrm{SPA}}(\beta, q_{0}) =
\frac{1}{2|k|} \,q_{0}^{2} - \frac{1}{\beta} \,\textrm{ln} \,z(\beta, q_{0})
\end{equation}
plays the role of an effective static free energy. It is not the free energy 
${\cal F}(\beta) = -T \ \textrm{ln} \,\mathcal{Z}(\beta)$ of the total
self-bound system,  
but the one of the constituents moving in a mean field that is kept fixed at 
the static collective variable $q_{0}$. For this reason we call 
$\mathcal{F}^{\textrm{SPA}}(\beta, q_{0})$ the ``intrinsic free energy'' 
from now on, where the index ``SPA'' already refers to the simplest 
``static path approximation'' to (\ref{Z-athaly}) [see Sec.~\ref{convPSPA}].
In (\ref{FSPA}) there appears the grand canonical partition function
\bal{zbeta}
z(\beta, q_{0}) & = & \textrm{Tr} \,\exp \left( -\beta (\hat{h}_{0}(q_{0})
- \mu \hat{A}) \right) \nonumber \\
& = & \prod_{l} \left( 1 + \exp \left( -\beta (\epsilon_{l}(q_{0}) - \mu) 
\right) \right)
\end{eqnarray}
belonging to the static part of the Hamiltonian (\ref{twobodham}) in mean 
field approximation
\bel{1bHam-stat} \hat{h}_{0}(q_{0}) = \hat{H} + q_{0} \,\hat{F}.
\end{equation}
Obviously, $ \hat{h}_{0}(q_{0})$ is a sum of only one body operators
with eigenenergies $\epsilon_{l}(q_{0})$.

All contributions from the dynamical part of the auxiliary path $q(\tau)$
are contained in the factor $C(\beta,q_0)$ which can be formally
written as \cite{ath.aly:npa:97}
\bel{corrfactor} 
C(\beta, q_{0}) = \int {\cal D}'q \ \exp
\left( -\frac{\beta}{|k|} \sum_{r>0} |q_{r}|^{2} + \textrm{ln} \langle
\hat{\mathcal{U}}_{q} \rangle_{q_{0}} \right)
\end{equation}
with the measure
\bel{measure}
{\cal D}'q = \lim_{\substack{N\rightarrow\infty \\
N\varepsilon = \hbar\beta}} \prod_{r=1}^{(N-1)/2}
\frac{\beta}{\pi |k|} \ d\textrm{Re}(q_{r}) \,d\textrm{Im}(q_{r}) \ .
\end{equation}
Here, the thermal expectation value
of the evolution operator $\hat{\mathcal{U}}_{q}$ has to be evaluated 
with respect to the static one-body Hamiltonian (\ref{1bHam-stat}) and 
reads in terms of a time-ordered product
\bel{meanUq} 
\langle \hat{\mathcal{U}}_{q} \rangle_{q_{0}} =
\left\langle \hat{\cal T} \exp \left( -\frac{1}{\hbar}
\int_{0}^{\hbar\beta} d\tau \,\hat{h}_{1}(\tau, q_{r}) \right)
\right\rangle_{q_{0}} \ .
\end{equation}
The ``dynamical'' Hamiltonian
\bel{1bHam-td-int}
\hat{h}_{1}(\tau, q_{r}) = \hat{F}(\tau) \,\delta q(\tau)
\end{equation}
with $\delta q(\tau) = q(\tau) - q_{0}$
may be understood as the time dependent correction to the static
mean field Hamiltonian $\hat h_0(q_0)$ given in
(\ref{1bHam-stat}). Hence, it is convenient to work  in an interaction picture
based on  $\hat h_0(q_0)$ and define time dependent operators 
as, e.g.
\bel{interact}
\hat{F}(\tau) = \expo^{\hat{h}_{0}(q_{0}) \tau/\hbar} \ \hat{F}
\ \expo^{-\hat{h}_{0}(q_{0}) \tau/\hbar} \ .
\end{equation}
The partition function (\ref{Z-athaly}) together with (\ref{FSPA}) and
(\ref{corrfactor}) is 
still exact, however, written in a way which allows for a systematic
approximative evaluation.
Namely, all  quantum fluctuations, i.e.\  modes with Matsubara
frequencies $\nu_r\neq 0$, are hidden in
$\langle\hat{\mathcal{U}}_{q}\rangle_{q_{0}}$. Thus, 
 the basic idea is to successively account for
dynamical information in ${\cal Z}(\beta)$ by expanding the thermal
expectation value 
$\langle\hat{\mathcal{U}}_{q}\rangle_{q_{0}}$ around its static value
in terms of the  deviations $\delta q(\tau)$. This way, generalizing
Eq.~(27) of \cite{ath.aly:npa:97} to fourth 
order in the  $q_{r}$ one finds:
\bal{logU} 
& & \textrm{ln} \,\langle\hat{\mathcal{U}}_{q}\rangle_{q_{0}}^{\textrm{ePSPA}} 
\nonumber \\
& = & \frac{1}{2! \,\hbar^{2}} \sum \int \ q_{r}q_{s}
\ \expo^{\imag\nu_{r}\tau_{r}} \expo^{\imag\nu_{s}\tau_{s}} 
\ \langle \hat{\cal T} \hat{F}(\tau_{r})\hat{F}(\tau_{s}) \rangle_{q_{0}} 
\nonumber \\
& + & \frac{1}{3! \,\hbar^{3}} \sum \int \ q_{r}q_{s}q_{t}
\ \expo^{\imag\nu_{r}\tau_{r}} \expo^{\imag\nu_{s}\tau_{s}} 
\expo^{\imag\nu_{t}\tau_{t}} \nonumber \\
& & \qquad\qquad\quad \times \ \langle \hat{\cal T} \hat{F}(\tau_{r})
\hat{F}(\tau_{s})\hat{F}(\tau_{t}) 
\rangle_{q_{0}} \nonumber \\
& + & \frac{1}{4! \,\hbar^{4}} \sum \int \ q_{r}q_{s}q_{t}q_{u}
\ \expo^{\imag\nu_{r}\tau_{r}} \expo^{\imag\nu_{s}\tau_{s}} 
\expo^{\imag\nu_{t}\tau_{t}} \expo^{\imag\nu_{u}\tau_{u}} \nonumber \\
& & \qquad\qquad\quad \times \ \langle \hat{\cal T} \hat{F}(\tau_{r})
\hat{F}(\tau_{s})\hat{F}(\tau_{t})\hat{F}(\tau_{u}) \rangle_{q_{0}} 
\nonumber \\
& - & \frac{1}{8 \,\hbar^{4}} \sum \int \ q_{r}q_{s}q_{t}q_{u}
\ \expo^{\imag\nu_{r}\tau_{r}} \expo^{\imag\nu_{s}\tau_{s}} 
\expo^{\imag\nu_{t}\tau_{t}} \expo^{\imag\nu_{u}\tau_{u}} \nonumber \\
& & \qquad\qquad\quad \times \ \langle\hat{\cal T} \hat{F}(\tau_{r})
\hat{F}(\tau_{s}) \rangle_{q_{0}} 
\langle \hat{\cal T} \hat{F}(\tau_{t})\hat{F}(\tau_{u}) \rangle_{q_{0}}
\nonumber \\
& + & {\cal O}(q_{r}^{5})
\end{eqnarray}
The symbol $\sum \int$ abbreviates summation over all involved 
$r, \,s, \,\ldots \not= 0$ and integration of all involved
$\tau_{r}, \,\tau_{s}, \,\ldots$ from $0$ to $\hbar\beta$.
Because of the $\tau$-integrations and the fact that
$\langle \hat{F}(\tau) \rangle_{q_{0}}$ is $\tau$-independent all terms
involving such a factor --- for instance terms linear in $q_r$ --- 
vanish immediately and are therefore omitted in  (\ref{logU}).

\subsection{Conventional PSPA -- Expansion to second order}
\label{convPSPA}

The simplest approximation to ${\cal Z}(\beta)$, coined
the static path approximation (SPA), neglects all dynamical
contributions, i.e. one puts $\hat{h}_{1}(\tau, q_{r})
\equiv 0$. As a result ${\cal C}^{\textrm{SPA}}(\beta, q_{0})
\equiv 1$. The SPA is the classical limit where all  many body
quantum fluctuations are absent. For lower temperatures when quantum
properties tend to become important, fluctuations around the classical
limit can be incorporated within the conventional version of the perturbed 
static path approximation (PSPA). There, the expansion (\ref{logU}) is 
truncated after the second order terms in the $q_{r}$ 
\cite{pug.bop.brr:ap:91,ath.aly:npa:97,ror.can:plb:97,ror.rip:npa:98,agb.ana:plb:98,agb.sas.dej.shs:prc:98,ruc.hoh:pre:01,ror.can.rip:ap:99,rossignoli}, 
which effectively means to describe quantum effects on the RPA level.
Within the PSPA the dynamical factor $C(\beta,q_0)$ is approximated by  
\cite{ruc.hoh:pre:01}
\bal{C-PSPA} 
& & C^{\textrm{PSPA}}(\beta, q_{0}) = \\
& & \int {\cal D}'q \ \exp \left( -\frac{\beta}{|k|} \sum_{r>0} 
\left( 1 + k\chi(q_{0}, \imag\nu_{r}) \right) q_{r} q_{-r} \right) \, .
\nonumber
\end{eqnarray}
Here, the FF-response function $\chi(q_{0}, \omega)$ defined by
\bel{defchi}
\delta \langle\hat{F}\rangle_{q_{0}}(\omega) = 
-\chi(q_{0}, \omega) \ \delta q(\omega)
\end{equation}
has to be evaluated at the Matsubara frequencies $\imag\nu_{r}$ along the
imaginary axis. Because of $q_{-r} = q_{r}^{*}$ all integrals in 
(\ref{C-PSPA}) are of Gaussian type, and cause no problem as long as
\bel{convcond}
\lambda_{r}(\beta, q_{0}) \equiv 1 + k\chi(q_{0}, \imag\nu_{r}) > 0 
\quad \textrm{for \ all} \quad r > 0 \ .
\end{equation}
For systems with $q_{0}$-regions where the local RPA frequencies 
$\omega_{\nu}(\beta, q_{0})$ obtained by the condition 
$1 + k\chi(q_{0}, \omega_{\nu}) = 0$
are purely imaginary eq.(\ref{convcond}) defines a 
condition for that temperature below which the conventional version of the 
PSPA breaks down due to a vanishing $\lambda_{1}(\beta, q_{0})$. 
This breakdown temperature $T_0$ where $\lambda_1(1/T_{0}, q_0) = 0$ for the 
first time is known from the theory of dissipative tunneling \cite{grabert} 
as the ``crossover temperature'' (for many body systems see
\cite{ruc.hoh:pre:01}) 
\bel{T0}
T_{0} =  \textrm{max}_{q_0} 
\ \frac{\hbar \,|\omega_{\nu}^{\text{inst}}(q_{0})|}{2\pi} 
\end{equation}
with $\omega_{\nu}^{\text{inst}}(q_{0})$ as imaginary local RPA frequencies. 
Now, the condition (\ref{convcond}) is guaranteed for all $T > T_0$ and we
obtain for the partition function within conventional PSPA
\cite{ath.aly:npa:97,ruc.hoh:pre:01}
the integral (\ref{Z-athaly}) with the dynamical factor $C(\beta, q_{0})$ 
providing improvement over pure SPA taken as
\bel{defCPSPA}
C^{\textrm{PSPA}}(\beta, q_{0}) = 
\prod_{r>0} \frac{1}{\lambda_{r}(\beta, q_{0})}\, .
\end{equation}

\subsection{Extended PSPA -- Expansion to fourth order}
\label{extPSPA}

The instabilities that lead to the breakdown of the conventional PSPA at 
$T_{0}$ are due to the fact that all fluctuations $q_{r}$ are treated only up 
to second order. Since the Gaussians in $q_{r}$ have a
typical width [see (\ref{C-PSPA})]
\bel{width}
\Delta_{r} = \sqrt{\frac{|k|}{2\beta\lambda_{r}(\beta, q_{0})}}
= \sqrt{\frac{|k| T}{2\lambda_{r}(\beta, q_{0})}}\ \ , 
\end{equation}
first $\Delta_1$ grows as temperature approaches $T_{0}$ from
above, i.e.\ $\lambda_{1}(\beta, q_{0})\to 0$. Correspondingly,
fluctuations in  
$q_{1}$-direction increase which in turn renders the Gaussian
approximation for the $q_1$-mode insufficient. In contrast, the other
amplitudes $q_{2}, \,q_{3}, 
\,\ldots$ remain still sufficiently small for a harmonic approximation
to be valid. Therefore in the expansion (\ref{logU}) terms 
higher than second order must be taken into account {\em only} for the $q_{1}$ 
and $q_{-1}$ mode. This is very similar to the procedure used in the 
framework of dissipative tunneling  
in order to overcome the irregularity of the decay rate at $T_0$
\cite{grabert}.  Working along these lines 
a consistent approximation to  the partition function (\ref{Z-athaly})
is obtained for temperatures near $T_0$ 
by taking for $C(\beta, q_{0})$ 
\bel{Z-4ord}
C^{\textrm{ePSPA}}(\beta, q_{0}) = 
\int {\cal D}'q \ \exp \left( -\frac{\beta}{|k|} \,A(\beta, q_{0}, q_{r}) 
\right).
\end{equation}
The relevant effective multidimensional fluctuation potential turns
out to be 
\bal{defA}
A(\beta, q_{0}, q_{r}) & = & 
\sum_{r>0} \lambda_{r}(\beta, q_{0}) \,q_{r} q_{-r} \\
& & + 3 c_{3}^{+}(\beta, q_{0}) \,q_{1}^{2}q_{-2} 
    + 3 c_{3}^{-}(\beta, q_{0}) \,q_{-1}^{2}q_{2} \nonumber \\
& & + 6 c_{4}(\beta, q_{0}) \,q_{1}^{2}q_{-1}^{2} \ . \nonumber
\end{eqnarray}
Here, other terms e.g.\ of the type $q_1 q_s q_{-s-1}$ or $q_1 q_{-1}
q_s q_{-s}$ ($s>1$) are negligible in the sense of a semiclassical
approximation as they contribute only in lower order \cite{grabert}. 
The relevant coefficients containing third and fourth order $\hat{F}$
correlations read: 
\begin{eqnarray}
& & c_{3}^{+}(\beta, q_{0}) = \label{defC3+} \\
& & \frac{-|k| / \beta}{3! \,\hbar^{3}} \int 
\ \expo^{\imag\nu_{+1}\tau_{r}} \expo^{\imag\nu_{+1}\tau_{s}} 
\expo^{\imag\nu_{-2}\tau_{t}} 
\ \langle \hat{\cal T} \hat{F}(\tau_{r})\hat{F}(\tau_{s})\hat{F}(\tau_{t}) 
\rangle_{q_{0}} \nonumber \\
& & c_{3}^{-}(\beta, q_{0}) = \label{defC3-} \\ 
& & \frac{-|k| / \beta}{3! \,\hbar^{3}} \int 
\ \expo^{\imag\nu_{-1}\tau_{r}} \expo^{\imag\nu_{-1}\tau_{s}} 
\expo^{\imag\nu_{+2}\tau_{t}} 
\ \langle \hat{\cal T} \hat{F}(\tau_{r})\hat{F}(\tau_{s})\hat{F}(\tau_{t}) 
\rangle_{q_{0}} \nonumber \\
& & c_{4}(\beta, q_{0}) = \label{defC4} \\
& & \frac{-|k| / \beta}{4! \,\hbar^{4}} \int 
\ \expo^{\imag\nu_{+1}\tau_{r}} \expo^{\imag\nu_{+1}\tau_{s}} 
\expo^{\imag\nu_{-1}\tau_{t}} \expo^{\imag\nu_{-1}\tau_{u}} 
\nonumber \\
& & \qquad\qquad \times \ \left( \langle \hat{\cal T} \hat{F}(\tau_{r})
\hat{F}(\tau_{s})\hat{F}(\tau_{t})\hat{F}(\tau_{u}) \rangle_{q_{0}} \right. 
\nonumber \\
& & \qquad\qquad\quad - \left. 3 \langle \hat{\cal T} \hat{F}(\tau_{r})
\hat{F}(\tau_{s}) \rangle_{q_{0}} 
\langle \hat{\cal T} \hat{F}(\tau_{t})\hat{F}(\tau_{u}) \rangle_{q_{0}}
\right). \nonumber
\end{eqnarray}
The integrals symbolize integrations over all involved 
$\tau$ from $0$ to $\hbar\beta$.
Mind that only Matsubara frequencies with indices 
$r, s, t, u = \pm 1, \pm 2$ enter these formulas.
The explicit evaluation of these coefficients is a
crucial step in order to apply the ePSPA to many body systems as we
will demonstrate in detail in Sec.~\ref{coeff}. In particular, we will
prove in Sec.~\ref{coeff} [see (\ref{sumrule})] that the  
$\tau$ integrations of (\ref{logU}) imply the sum rule 
$r + s + t + \ldots = 0$ for the indices involved which considerably
simplifies the calculation.  

Let us now turn to the $q_r$ integrals in (\ref{Z-4ord}). Integrals
with $|r| > 1$ are still of Gaussian  
type and can easily be carried out as long as $\lambda_2(\beta,q_0)>0$
is sufficiently large. While all integrations for $|r| > 2$ just provide
factors $1/\lambda_r$ like in conventional PSPA, the integrals over 
$q_{2}$ and $q_{-2} = q_{2}^{*}$ require special care. 
To this end the variables $q_{2}$ and $q_{-2}$ are transformed to 
variables $q_{2}' = \textrm{Re} \,q_{2}$ and $q_{2}'' = \textrm{Im} \,q_{2}$ 
which cast the relevant part of (\ref{defA}) in the form
\bal{q2exp}
a(\beta,q_0,q_2',q_2'') & = & 
\lambda_{2}(\beta, q_{0}) \,(q_{2}')^{2} \\
& + & 3 \left( c_{3}^{-}(\beta, q_{0}) \,q_{-1}^{2} + c_{3}^{+}(\beta, q_{0}) 
\,q_{1}^{2} \right) \,q_{2}' \nonumber \\
& + & \lambda_{2}(\beta, q_{0}) \,(q_{2}'')^{2} \nonumber \\
& + & 3\imag \left( c_{3}^{-}(\beta, q_{0}) \,q_{-1}^{2} - c_{3}^{+}(\beta, 
q_{0}) \,q_{1}^{2} \right) \,q_{2}'' \ . \nonumber
\end{eqnarray}
The corresponding two 
Gaussian integrals can be easily solved:
\bal{q2Gauss}
& & \frac{\beta}{\pi |k|} \,\int dq_2' dq_2'' \ \exp \left(-\frac{\beta}{|k|}
\ a(\beta,q_0,q_2',q_2'')\right) \\
& = & \frac{1}{\lambda_{2}(\beta, q_{0})} \ \exp \left( \frac{\beta}{|k|} 
\ \frac{9 c_{3}^{-}(\beta, q_{0}) c_{3}^{+}(\beta, q_{0})}
{\lambda_{2}(\beta, q_{0})} \ q_{1}^{2}q_{-1}^{2} \right) \ . \nonumber
\end{eqnarray}

What is left are the $q_{1}$-integrals. These are integrals over an
exponential with an exponent given by
\bal{q1exp}
\lambda_{1}(\beta, q_{0}) \,q_{1}q_{-1} + B(\beta, q_{0}) 
\,q_{1}^{2}q_{-1}^{2} \ ,
\end{eqnarray}
where the coefficient $B(\beta, q_{0})$ is defined as
\bel{defB}
B(\beta, q_{0}) = 6 c_{4}(\beta, q_{0}) - \frac{9 c_{3}^{-}(\beta, q_{0}) 
c_{3}^{+}(\beta, q_{0})}{\lambda_{2}(\beta, q_{0})} \ .
\end{equation}
To proceed further it is convenient to 
introduce polar coordinates $\rho$ and $\phi$ instead of $q_{1}$ and 
$q_{-1} = q_{1}^{*}$, such that the 
phase integral can be done. After an additional substitution $z = \rho^{2}$ 
the $q_1$ integrals reduce to
\bal{q1int}
& & \frac{\beta}{\pi |k|} \int_{0}^{2\pi} \!\!d\phi \int_{0}^{\infty}
\!\!d\rho \ \rho \ \exp \left( -\frac{\beta}{|k|} \left( \lambda_{1} 
\,\rho^{2} + B \,\rho^{4} \right) \right) \\
& = & \frac{\beta}{|k|} \int_{0}^{\infty} \!\!dz \ \exp 
\left( -\frac{\beta}{|k|} \left( \lambda_{1}(\beta, q_{0}) \,z 
+ B(\beta, q_{0}) \,z^{2} \right) \right) \ . \nonumber
\end{eqnarray}

Eventually,  together with the contributions from (\ref{q2Gauss}) and 
(\ref{q1int}) the expression (\ref{Z-4ord}) for the 
new dynamical factor reads in "extended PSPA"
\bal{C-extPSPA}
& & C^{\textrm{ePSPA}}(\beta, q_{0}) = 
\prod_{r>1} \frac{1}{\lambda_{r}(\beta, q_{0})} \\
& \times & \frac{\beta}{|k|} \int_{0}^{\infty} dz 
\,\exp \left( -\frac{\beta}{|k|} 
\left( \lambda_{1}(\beta, q_{0}) z + B(\beta, q_{0}) z^{2} \right) \right) \ .
\nonumber
\end{eqnarray}
In case that $q_1$ fluctuations are small so that third and fourth
order terms in $q_{1}$ and $q_{-1}$ are negligible 
--- $B(\beta, q_{0}) \ll \lambda_{1}(\beta, q_{0})$ --- 
the factor $C^{\textrm{ePSPA}}(\beta, q_{0})$ reduces to 
$C^{\textrm{PSPA}}(\beta, q_{0})$ [see (\ref{defCPSPA})]. For
temperatures $T<T_0$ and $q_0$ for which $\lambda_1 < 0$ but
$|\lambda_1|$ again sufficiently large, 
the integral in (\ref{C-extPSPA}) can be evaluated in stationary
phase approximation. The result is
\bal{ePSPAlow}
C^{\textrm{ePSPA}}(\beta, q_{0}) & = &  
\prod_{r>1} \frac{1}{\lambda_{r}(\beta, q_{0})} \\
& \times & \sqrt{\frac{2\pi \beta}{|k| \,B(\beta,q_0)}}
\ \exp\left(\frac{\beta}{|k|}\frac{\lambda_1^2(\beta,q_0)}
{4 B(\beta,q_0)}\right) \ . \nonumber
\end{eqnarray}
Here, the exponential reveals the change of stability at $T=T_0$ as
now the contribution to ${\cal Z}(\beta)$ of those $q_0$ where
$\lambda_1<0$ is enhanced compared to their static value $\exp[-\beta 
{\cal F}^{\textrm{SPA}}(\beta,q_0)]$.

The ePSPA smoothly connects the temperature range above $T_0$ with
that below $T_0$. Its precise lower bound of
validity is determined by two conditions: $B(\beta, q_{0})$ defined in
(\ref{defB}) and $\lambda_2(\beta, q_{0})$ [see (\ref{convcond})] 
must both be positive and sufficiently large
{\em for all} $q_{0}$. Even if $B>0$ also for $T<T_0$  
 the ePSPA definitely fails at $T=T_0/2$ where
 $\lambda_{2}$ vanishes. Physically, the instability of
the PSPA at $T = T_0$ corresponds to an instability in $q_{1}$ direction 
in functional space of that classical mean field solution $q_0 = q_0^c$ 
for which $\lambda_1(1/T_{0}, q_0^c) = 0$. 
At the temperature where the ePSPA breaks down, i.e.\ where $\lambda_2=0$, 
the classical field at $q_0^c$ becomes unstable in $q_{2}$ direction 
in functional space. This scenario proceeds with decreasing temperature.

\subsection{Treatment of low temperatures $T\ll T_0$}
\label{lowtemp}

In the last section we have proposed an extension of the conventional PSPA
that is applicable down to temperatures not too far below $T_{0}$. As
already mentioned, this approximation breaks down 
near $\lambda_2=0$ where the $q_2$-mode amplitude becomes large. Naively, one
could think to ``regularize'' the $q_2$-mode divergency in the 
same way as for the $q_1$-mode around $T_0$. However, if both $q_1$
{\em and} $q_2$ are large even higher order coupling terms to $q_r$ modes 
($|r| > 2$) in the expansion must be taken into account
[compare our remark after eq. (\ref{defA})]. 
This proceeds to the divergencies at $\lambda_3=0$ and so on. Hence, 
practically an analytical treatment analogous to the $T_0$ case is no 
longer possible for temperatures $T\ll T_0$. In particular, 
for $T\to 0$ {\em all} modes $q_r$ must be assumed to be 
large so that in the worst case any kind of semiclassical
approximation to ${\cal Z}(\beta)$ fails. 

On the other hand, for $T\ll T_0$, but still $T>0$, we expect that
integrals of the type (\ref{Z-athaly}) are dominated by the
contributions around the minima
$q_{0,i}^{min}, i=1,\ldots, M$ of  
the intrinsic free energy (\ref{FSPA}) provided 
$\beta {\cal F}^{\textrm{SPA}}(\beta, q_0^c)$ exceeds 
$\beta {\cal F}^{\textrm{SPA}}(\beta, q_{0,i}^{min})$ by terms
sufficiently larger than of 
order 1. Under the additional condition that the    
correction factor $C^{\textrm{ePSPA}}(\beta, q_{0})$ is a 
smooth function of $q_{0}$ compared to the exponential we can apply a saddle 
point approximation to evaluate (\ref{Z-athaly}) together with 
(\ref{C-extPSPA}):
\bal{Z-extPSPA-saddle}
\mathcal{Z}^{\textrm{ePSPA}}(\beta) & \approx & 
\sum_{i=1}^M \frac{1}{\sqrt{|k| \,d^{2} \mathcal{F}^{\textrm{SPA}} 
/ dq_{0}^{2} |_{q_{0,i}^{min}}}} \\
& \times & \exp \left[-\beta \mathcal{F}^{\textrm{SPA}}(\beta, q_{0,i}^{min})
\right] \,C^{\textrm{ePSPA}}(\beta, q_{0,i}^{min}) \ . \nonumber 
\end{eqnarray}
The {\em assumption} that $C^{\textrm{ePSPA}}(\beta, q_{0})$ is smooth 
compared to $\exp [-\beta\mathcal{F}^{\textrm{SPA}}(\beta, q_{0})]$ 
--- variations of the dynamical factor are negligible on the scale on which 
the exponential varies --- bases on the fact that all approximations of SPA 
type are semiclassical approximations. This condition is certainly  
not fulfilled in regions where the dynamical factor diverges. 
The divergencies, 0 whenever $\lambda_r(\beta,q_0)=0$ ($r>1$)
only appear in those $q_0$-regions where the RPA frequencies are imaginary 
and of sufficiently large amount, thus, corresponding to saddle points of 
the full free energy. Therefore we may assume that they need not be 
considered for low temperatures (but still sufficiently above $T=0$). 
Accordingly, the following interpolation between the ePSPA and the
saddle point approximation (\ref{Z-extPSPA-saddle}) seems to be natural:
For $T\ll T_0$ we restrict the $q_0$ integration in (\ref{Z-athaly}) with 
(\ref{C-extPSPA}) to those
regions where $\lambda_1>0$ meaning that {\em all} $\lambda_r>0$, i.e.,
\bel{C-LTA}
C^{\textrm{LTA}}(\beta, q_{0}) = \theta[\lambda_1(\beta,q_0)]\
C^{\textrm{ePSPA}}(\beta, q_{0}) 
\end{equation}
where $\theta(\cdot)$ denotes the step function.
This interpolation is called ``low temperature approximation'' (LTA) 
henceforth. It requires to solve $\lambda_1(\beta,q_0^c)=0$ for given
$\beta > \beta_0 = 1/T_{0}$ to determine $q_0^c(\beta)$. As a
continuous function of $\beta$ starting at $q_0^c(\beta_0)$,
$q_0^c(\beta)$ describes the 
broadening of the instability region around $q_0^c(\beta_0)$ where
$\lambda_1<0$ \cite{anj.grh:pa:92}.
The ePSPA extends to temperatures somewhat below 
$T_0$, while the LTA is valid if with decreasing
temperature $T<T_0$ the contribution of the instability region around
$q_0^c(\beta_0)$  to the $q_0$ integral becomes negligible. Therefore 
both approximations match in a narrow temperature range below $T_0$.
Further, by formally extending the LTA to $T > T_{0}$ one recovers 
the ePSPA since then $\lambda_1>0$ for all $q_0$, while for $T \ll T_{0}$ 
the saddle point result 
(\ref{Z-extPSPA-saddle}) is regained. As already mentioned, in general
there is a lower bound for the LTA. Contributions of
mean fields around $q_0^c$ can only be neglected as long as the
exponential enhancement in
(\ref{ePSPAlow}) due to the dynamical instability is still compensated
for by the  
difference between ${\cal F}^{\textrm{SPA}}(\beta, q_0^c)$ and 
${\cal F}^{\textrm{SPA}}(\beta, q_{0,i}^{min})$. For very low temperatures 
this can no longer be taken for granted.

We want to remark that for specific systems further 
approximations may exist in the region $T\ll T_0$, possibly  better
adapted to the problem in 
question. This may be particularly true for very low temperatures
$T\to 0$ if new phenomena, sometimes associated with the appearance of
new symmetries like  in the case of  Goldstone modes, require special
care (cf. the cases of a one-dimensional double well and the 
Lipkin-Meshkov-Glick model  below in Sects.~\ref{1dim} and \ref{manybody}).
As long as one is interested in approximation schemes applicable to general
situations without referring to individual system properties,
however, the LTA presented above seems to us the only consistent one.

\section{Application I: One dimensional potential}
\label{1dim}

In order to illustrate
the utility of the 
the above approximations we turn to 
 exactly solvable models and first
apply the formalism to the one-dimensional case. 

Consider a particle of mass 
$M$  in a one dimensional potential $V(q)$. The imaginary time path
integral representation is given by
\bel{Z-pathint}
\mathcal{Z}(\beta) = \int dq' \oint_{q'} {\cal D}q(\tau) 
\ \exp \left( -S^{\textrm{E}}
[q(\tau), \dot{q}(\tau)] / \hbar \right)
\end{equation}
with the Euclidean action
\bel{S-Euklid}
S^{\textrm{E}}[q(\tau), \dot{q}(\tau)] = \int_{0}^{\hbar\beta} d\tau 
\left( \frac{M}{2} \,\dot{q}^{2}(\tau) + V(q(\tau)) \right) \ .
\end{equation}
In (\ref{Z-pathint}) we first have to sum over all paths $q(\tau)$
starting and ending at a given end-point $q(0) = q(\hbar\beta) = q'$ 
and second sum up contributions from all these end-points $q'$.

In order to be able to apply the ideas of Sec.~\ref{extPSPA}, the
potential  
must be expanded up to fourth order in the Fourier coefficients $q_{r}$ 
($|r| > 0$) introduced in (\ref{fluctcoord}). After carrying out the $\tau$ 
integration we obtain for the exponent:
\bal{exponent}
-S^{\textrm{E}}[q_{r}] / \hbar & = & -\beta V(q_{0})
- \beta \sum_{r>0} \left( M\nu_{r}^{2} + V''(q_{0}) \right) |q_{r}|^{2} 
\nonumber \\
& & - \frac{\beta V^{(3)}(q_{0})}{3!} \,\sum_{r,s,t \ne 0} 
\delta_{r+s+t, 0} \ q_{r} q_{s} q_{t} \\
& & - \frac{\beta V^{(4)}(q_{0})}{4!} \,\sum_{r,s,t,u \ne 0} 
\delta_{r+s+t+u, 0} \ q_{r} q_{s} q_{t} q_{u}\, . \nonumber 
\end{eqnarray}
For the case of one dimensional potentials the sum rule $r + s + t + u = 0$ 
simply stems from this $\tau$ integration. Here $V(q_{0})$ plays the same 
role as $\mathcal{F}^{\textrm{SPA}}(\beta, q_{0})$ in 
(\ref{Z-athaly}).
As before the coefficients $q_{1}$ and $q_{-1}$ become large at the crossover 
temperature 
\bel{T01D}
T_{0} = \textrm{max}_{q_{0}} \ \frac{\hbar}{2\pi} \sqrt{\frac{-V''(q_{0})}{M}}
\end{equation}
and therefore have to be taken into account up to fourth order. Then, with 
$\lambda_{r}(\beta, q_{0}) \equiv M\nu_{r}^{2} + V''(q_{0})$, 
$c_{3}^{+}(\beta, q_{0}) = c_{3}^{-}(\beta, q_{0}) \equiv V^{(3)}(q_{0}) / 3!$,
and $c_{4}(\beta, q_{0}) \equiv V^{(4)}(q_{0}) / 4!$ we gain
\bel{1DZextPSPA}
\mathcal{Z}(\beta) = 
\sqrt{\frac{M}{2\pi \hbar^{2}\beta}} \int dq_{0} 
\ \exp \left[ -\beta V(q_{0}) \right] \ C(\beta, q_{0}) \ .
\end{equation}
In case of the pure SPA we have $C^{\textrm{SPA}}(\beta, q_{0}) 
\equiv 1$, in case of the conventional PSPA
\bel{1dC-PSPA}
C^{\textrm{PSPA}}(\beta, q_{0}) = 
\prod_{r>0} \frac{M\nu_{r}^{2}}{M\nu_{r}^{2} + V''(q_{0})}\, ,
\end{equation}
and for the extended PSPA
\bal{1dC-extPSPA}
& & C^{\textrm{ePSPA}}(\beta, q_{0}) = 
\prod_{r>1} \frac{M\nu_{r}^{2}}{M\nu_{r}^{2} + V''(q_{0})} \times \\
& & \int_{0}^{\infty} dz \ \exp \Bigg\{ -\beta \Bigg[ \left( M\nu_{1}^{2} 
+ V''(q_{0}) \right) z \Bigg. \Bigg. \nonumber \\
& & \qquad\qquad\qquad + \Bigg. \Bigg. \frac{1}{4} \left( V^{(4)}(q_{0}) - 
\frac{(V^{(3)}(q_{0}))^{2}}{M\nu_{2}^{2} + V''(q_{0})} \right) z^{2} 
\Bigg] \Bigg\} \ . \nonumber
\end{eqnarray}

\begin{figure}
\resizebox{0.45\textwidth}{!}{\includegraphics{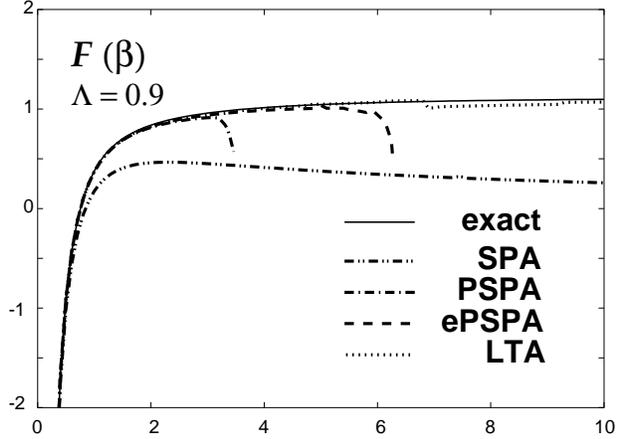}}
\caption{\label{fig-Fbeta1D09} 
${\cal F}(\beta)$ of the double well (\ref{doublewell}) computed for 
various approximations and $\Lambda = 0.9$.}
\end{figure}

Let us now turn to the specific case of a
particle of mass $M = 1/2$ ($\hbar = 1$)  moving in the quartic potential
\bel{doublewell}
V(q) = -\Lambda \,q^{2} + (1 - \Lambda) \,q^{4} + \frac{\Lambda^{2}}
{4 (1 - \Lambda)} 
\quad \textrm{with} \quad 0 \le \Lambda < 1 \ .
\end{equation}
This potential coincides with a quartic oscillator for $\Lambda = 0$ and
develops a barrier of height $V(0) = \Lambda^{2} / (4 (1 - \Lambda))$ 
for positive $\Lambda$. Potentials of the form 
(\ref{doublewell}) have often been used to qualitatively understand 
characteristic properties of high dimensional systems as e.g.\ phase
 transitions. Here we find that the shape of $V(q)$ describes the main
 features encoded in the intrinsic free energy of the Lipkin-Meshkov-Glick 
model which we will consider in detail below (see
 Sec.~\ref{appl_lip}). There, the appearance of a phase transition is
 equivalent to the development of a barrier for $\Lambda>0$ in $V(q)$.
Further, from (\ref{T01D}) and (\ref{doublewell}) the inverse crossover 
temperature can simply be derived as  
$\beta_{0} = 1 / T_{0} = 2\pi / \hbar \cdot \sqrt{M / (2\Lambda)} = 
\pi / (\hbar \sqrt{\Lambda})$. 
In fig.~\ref{fig-Fbeta1D09}  the free energy 
\bel{FfromZ}
{\cal F}(\beta) = -\frac{1}{\beta} \ \textrm{ln} \,\mathcal{Z}(\beta)
\end{equation}
is depicted within various approximations for $\Lambda = 0.9$.

\section{Application II: The Lipkin-Meshkov-Glick model}
\label{manybody}

The approximations derived in the preceeding sections are designed to
describe low temperature thermodynamic properties of interacting many
body systems. In contrast to simple one-dimensional cases, here, an
additional challenge arises through the calculation of the expansion
coefficients in (\ref{logU}) and (\ref{defC3+})--(\ref{defC4}).

\subsection{Expansion coefficients for many body systems}
\label{coeff}

For the Hamiltonian (\ref{twobodham}) the coefficients $c_{3}^{\pm}$ and 
$c_{4}$ are calculated by evaluating the
$\tau$-integrals in (\ref{defC3+})--(\ref{defC4}). The calculation is
performed completely parallel to the one presented in
\cite{ath.aly:npa:97}. Hence, we restrict ourselves here to
{\em one} contribution to the fourth order coefficient to exemplify 
the general strategy. We also will prove the sum rule $r + s + t + u = 0$ 
exploited in (\ref{defA})--(\ref{defC4}) here. To this end we consider 
general indices $r, \,s, \,t, \,u \ne 0$ at first. One has to rewrite 
the operators  $\hat{F}(\tau)$ in terms of creation and annihilation operators
$\hat{F}(\tau) = \sum_{kj} F_{kj} \hat{a}_{k}^{\dagger}(\tau)\hat{a}_{j}(\tau)$
and then calculate the time ordered averages using the finite temperature 
Wick theorem \cite{negelej.orlandh}. Evaluating all possible contractions 
the unperturbed temperature Green functions enter, i.e.\ 
\bal{Green}
-\langle \hat{\cal T} \hat{a}_{j}(\tau_{r}) \hat{a}_{k}^{\dagger}(\tau_{s})
\rangle_{q_{0}} & = & \delta_{jk} \ g_{k}^{(0)}(\tau_{r} - \tau_{s}) \\
& = & \delta_{jk} \,\frac{1}{\beta} \sum_{K = -\infty}^{\infty} 
\frac{\expo^{-\imag\omega_{K}(\tau_{r}-\tau_{s})}}{\imag\hbar\omega_{K} 
- e_{k}(q_{0})} \ , \nonumber 
\end{eqnarray}
where
$\omega_{K} = (2K + 1) \pi / \hbar\beta$  
(\footnote{Mind that our $\omega_{K}$ correspond to the $\nu_{k}$ of 
\cite{ath.aly:npa:97} and our $\nu_{r}$ are equivalent to the $\omega_{r}$ 
of \cite{ath.aly:npa:97}.})
and 
$e_{k}(q_{0})$ corresponds to the eigenenergies of (\ref{1bHam-stat}) via 
$e_{k}(q_{0}) = \epsilon_{k}(q_{0}) - \mu$. The integrands of all multiple 
$\tau$ integrals therefore turn into factors like e.g.
\bel{tauint}
\exp \left( \imag \,\frac{2\pi}{\hbar\beta} \,(r + I - O) \,\tau_{r} \right) 
\ .
\end{equation}
After carrying out all $\tau$ integrals we obtain the conditions
$r + I - O = 0$, $s - I + K = 0$, $t - K + M = 0$ and $u + O - M = 0$.
Putting these conditions together we easily obtain the requirement
\bel{sumrule}
r + s + t + u = 0 \ . 
\end{equation}
Moreover, we have only one summation index 
$I = -\infty \ldots \infty$ left besides $r, \,s, \,t \ne 0$
which will actually be further restricted by the requirement 
$r, \,s, \,t, \,u = \pm 1, \pm 2$ in (\ref{defC3+}) -- (\ref{defC4}).
One can show after some straightforward but lengthy algebra that 
{\em one} typical contribution to the fourth order coefficient 
$c_{4}(\beta, q_{0})$ reads:
\bal{coeff4-a}
& & \frac{-|k| / \beta}{4!} 
\sum_{i,k,m,o} F_{io}(q_{0}) F_{ki}(q_{0}) F_{mk}(q_{0}) F_{om}(q_{0}) 
\nonumber \\
& & \times \sum_{I = -\infty}^{\infty} 
\frac{1}{\imag\hbar\omega_{I} - e_{i}(q_{0})} 
\,\frac{1}{\imag\hbar (\omega_{I} + \nu_{r}) - e_{o}(q_{0})} \\
& & \quad\qquad \frac{1}{\imag\hbar (\omega_{I} - \nu_{s}) - e_{k}(q_{0})} 
\,\frac{1}{\imag\hbar (\omega_{I} - \nu_{s+t}) - e_{m}(q_{0})} \nonumber
\end{eqnarray}
The infinite sum over $I$ is calculated by exploiting the frequency summation 
technique \cite{fettera.waleckaj}. One uses the replacement 
$\imag\omega_{I} \to z$ and the residue theorem for 
contour integration in the complex $z$-plane together with the observation 
that the function $-\hbar\beta / (\exp(\hbar\beta z) + 1)$ has poles at 
$\imag\omega_{I}$ with residues $1$. After a deformation of the integration 
contour such that it encycles the poles of (\ref{coeff4-a})
we obtain the following final form:
\bal{coeff4-b}
& & \frac{|k|}{4!} 
\sum_{i,k,m,o} F_{io}(q_{0}) F_{ki}(q_{0}) F_{mk}(q_{0}) F_{om}(q_{0}) \\
& \times & \left\{ n(\epsilon_{i}) 
\,\frac{1}{\epsilon_{io} + \imag\hbar\nu_{r}}
\,\frac{1}{\epsilon_{ik} - \imag\hbar\nu_{s}}
\,\frac{1}{\epsilon_{im} - \imag\hbar\nu_{s+t}} \right. \nonumber \\
& & + \ n(\epsilon_{o}) 
\,\frac{1}{\epsilon_{oi} - \imag\hbar\nu_{r}}
\,\frac{1}{\epsilon_{ok} - \imag\hbar\nu_{r+s}}
\,\frac{1}{\epsilon_{om} - \imag\hbar\nu_{r+s+t}} \nonumber \\
& & + \ n(\epsilon_{k}) 
\,\frac{1}{\epsilon_{ki} + \imag\hbar\nu_{s}}
\,\frac{1}{\epsilon_{ko} + \imag\hbar\nu_{r+s}}
\,\frac{1}{\epsilon_{km} - \imag\hbar\nu_{t}} \nonumber \\
& & \left. + \ n(\epsilon_{m}) 
\,\frac{1}{\epsilon_{mi} + \imag\hbar\nu_{s+t}}
\,\frac{1}{\epsilon_{mo} + \imag\hbar\nu_{r+s+t}}
\,\frac{1}{\epsilon_{mk} + \imag\hbar\nu_{t}} \right\} \nonumber
\end{eqnarray}
Here $n(\epsilon) = (1 + \exp (\beta (\epsilon - \mu)))^{-1}$ are the 
Fermi occupation numbers and $\epsilon_{ik}(q_{0})$ are the energy 
differences between the states $|i(q_{0})\rangle$ and $|k(q_{0})\rangle$ of 
the static Hamiltonian $\hat{h}_{0}(q_0)$. This way, by solving the static 
one body Schr\"odinger equation belonging to (\ref{1bHam-stat}) 
and evaluating sums as in (\ref{coeff4-b}) all needed coefficients are 
known and the corresponding approximations to ${\cal Z}(\beta)$ can be 
applied. 

We note that the factor in brackets in (\ref{coeff4-b}) also appears 
in \cite{ath.aly:npa:97} where strength functions are calculated in 
the conventional PSPA formalism. There, time ordered expectation
values of products of one body operators 
$\hat{D} = \sum_{kj} D_{kj} \hat{a}_{k}^{\dagger} \hat{a}_{j}$  like  
$\langle \hat{\cal T} \hat{\cal U}_{q} \hat{D}^{\dagger}(\tau) \hat{D}(0) 
\rangle_{q_{0}}$ are evaluated. To this end $\hat{\cal U}_{q}$ [see
(\ref{meanUq}) and (\ref{logU})] have to be expanded up to 
second order in the $q_{r}$. As a consequence, fourth order terms in  
$\hat{a}_{k}^{\dagger}(\tau)\hat{a}_{j}(\tau)$ come into play even
though the approximation to  $\hat{\cal U}_{q}$ is still of Gaussian type.

\subsection{The model}
\label{appl_lip}

There are only few many body systems which allow for an exact
evaluation. One is the Lipkin-Meshkov-Glick model
\cite{lih.men.gla:np:65} which has been 
used to test the results of conventional PSPA for years
(see e.g.\ \cite{pug.bop.brr:ap:91,ath.aly:npa:97,ror.rip:npa:98}). 
In the last part of the paper we want to  
apply the approximations derived in Sec.~\ref{partfunc} to this
model.  The Lipkin-Meshkov-Glick 
Hamiltonian reads
\bel{lip_basHam}
\hat{\cal H} = 2\epsilon \hat{J}_{z} + 2k \hat{J}_{x}^{2} \ ,
\end{equation}
and has the structure of  a Hamiltonian with two body interaction with
 $\hat{H} = 2\epsilon \hat{J}_{z}$ and  
$\hat{F} = 2\hat{J}_{x}$. In (\ref{lip_basHam}) the operators $\hat{J}_{x}$ 
and $\hat{J}_{z}$ obey angular momentum commutation relations and $k = -|k|$ 
is a negative coupling constant describing an attractive 
interaction. 

Let us briefly recall \cite{pug.bop.brr:ap:91,ath.aly:npa:97,ror.rip:npa:98} 
the most important features of the PSPA applied to the Hamiltonian 
(\ref{lip_basHam}). The static part (\ref{1bHam-stat}) 
of the one body Hamiltonian is given by
\bel{lip_1bHam-stat}
\hat{h}_{0}(q_{0}) = 2\epsilon\hat{J}_{z} + 2q_{0}\hat{J}_{x}
\end{equation}
and its eigenvalues are $g$-fold degenerated
\bel{lip_epsbar}
\bar{\epsilon}^{2}(q_{0}) = \epsilon^{2} + q_{0}^{2} \ .
\end{equation}
Energy differences $\epsilon_{ik}(q_{0})$ can therefore only have the 
three different values $0, \pm 2\bar{\epsilon}(q_{0})$. 
For the grand canonical partition function belonging to 
(\ref{lip_1bHam-stat}) one easily obtains at a given collective coordinate  
$q_{0}$ [see (\ref{zbeta})] 
\bel{lip_zbeta}
z(\beta, q_{0}) = \left( 2 \cosh \left( \frac{\beta\bar{\epsilon}(q_{0})}{2} 
\right) \right)^{2g} \ .
\end{equation}
The PSPA correction factor (\ref{defCPSPA}) can be written as
\bel{lip_C-PSPA}
C^{\textrm{PSPA}}(\beta, q_{0}) = \frac{\sinh (\beta\bar{\epsilon}
(q_{0}))}{\beta\bar{\epsilon}(q_{0})} \ \frac{\hbar\beta\varpi(\beta, 
q_{0})/2}{\sinh (\hbar\beta\varpi(\beta, q_{0})/2)} \ 
\end{equation}
where the RPA frequencies read
\bel{lip_RPA}
\left( \frac{\hbar\varpi(\beta, q_{0})}{2} \right)^{2} = \bar{\epsilon}^{2}
(q_{0}) - \kappa \,\frac{\epsilon^{3}}{\bar{\epsilon}(q_{0})} 
\,\tanh \left( \frac{\beta\bar{\epsilon}(q_{0})}{2} \right) \ ,
\end{equation}
with the dimensionless coupling parameter%
(\footnote{Our $\kappa$ is defined as in \cite{ath.aly:npa:97} and 
should not be mixed up with the $\kappa = 1/k$ often used in nuclear 
physics \cite{bohra.mottelsonb.2}.})
\bel{defkappa}
\kappa = \frac{|k| g}{\epsilon} > 0.
\end{equation}
For $\kappa > 1$ the equilibrium 
value of $q_{0}$ undergoes a phase transition at a critical temperature 
$T_{c}$ which is determined from the condition 
that one can find a $q_{0}$ with zero RPA frequency from
(\ref{lip_RPA}). 
This phase transition manifests itself in the development of a barrier 
in the intrinsic free energy for $T < T_{c}$ corresponding to
imaginary RPA frequencies. 
Thus, the crossover temperature $T_{0}$ where  the factor
(\ref{lip_C-PSPA}) diverges is
determined from the condition $(\hbar\beta\varpi(\beta,
q_{0})/2)^{2}=-\pi^{2}$. We note that one always has $T_{0} < T_{c}$.
For $\kappa < 1$ no phase transition occurs and the RPA frequencies
are always real.

\subsection{Free energy}
\label{freeen}

For the Lipkin-Meshkov-Glick model the free energy has been
calculated in \cite{ath.aly:npa:97} within the conventional PSPA.
Here,  we want to
demonstrate the results of the extensions proposed  
above using the same set of parameters as in
\cite{pug.bop.brr:ap:91}, namely, 
\bel{paramPBB}
\epsilon = 5 \,\textrm{MeV} \quad \textrm{and} \quad g = 10 
\end{equation}
and various values for $\kappa$.

\begin{figure}
\resizebox{0.45\textwidth}{!}{\includegraphics{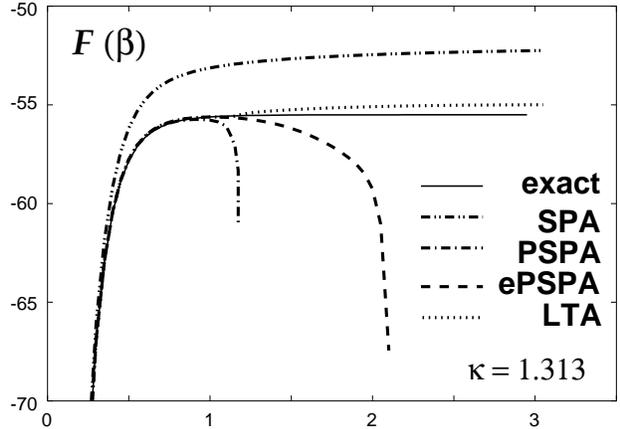}}
\caption{\label{fig-Fbeta1313enl} 
${\cal F}(\beta)$ of the Lipkin-Meshkov-Glick model (\ref{lip_basHam}) 
within various approximations for the parameters 
(\ref{paramPBB}) and $\kappa = 1.313$. The phase transition occurs at 
$\beta_{c} = 0.4 \,\textrm{MeV}^{-1}$ and the inverse crossover temperature 
is $\beta_{0} = 1.1392 \,\textrm{MeV}^{-1}$.}
\end{figure}

\begin{figure*}
\resizebox{0.95\textwidth}{!}{
\includegraphics[width=0.3\linewidth]{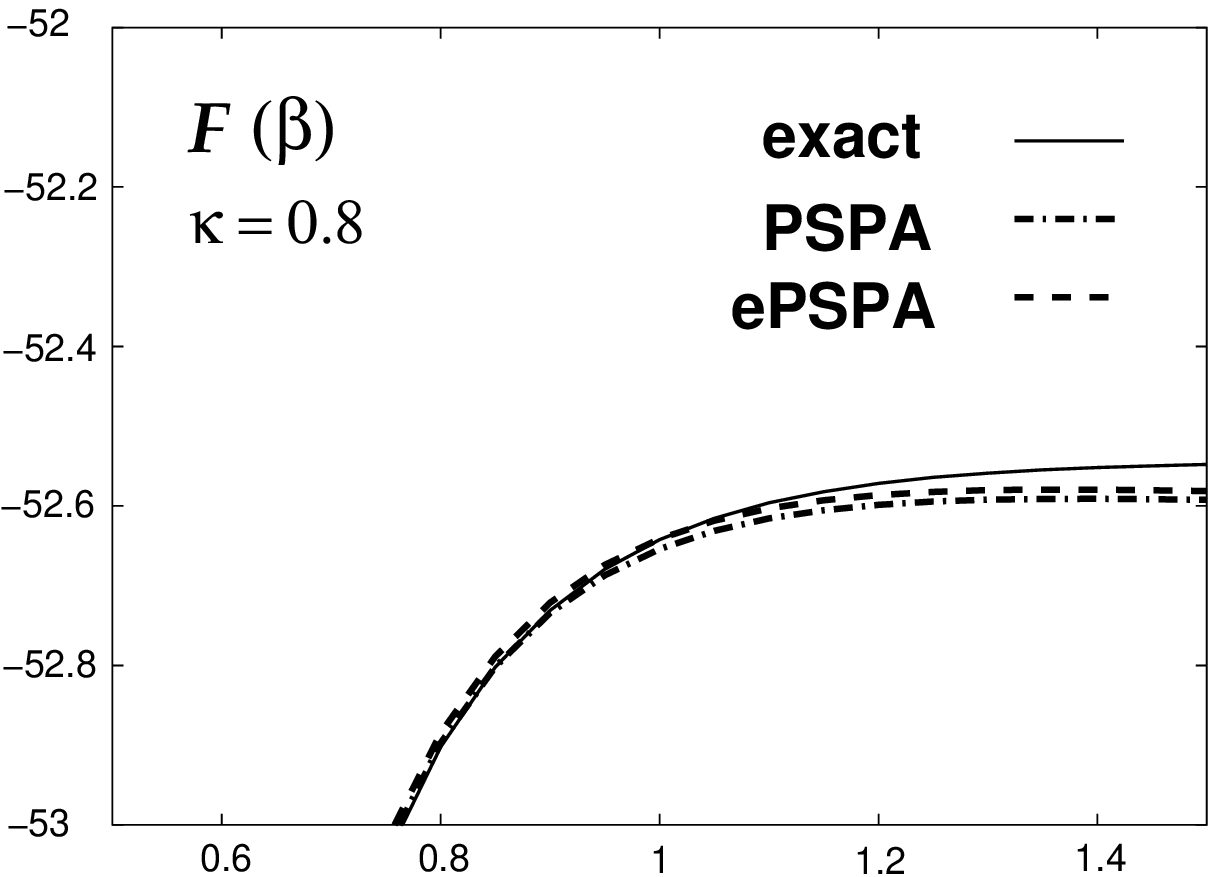}
\includegraphics[width=0.3\linewidth]{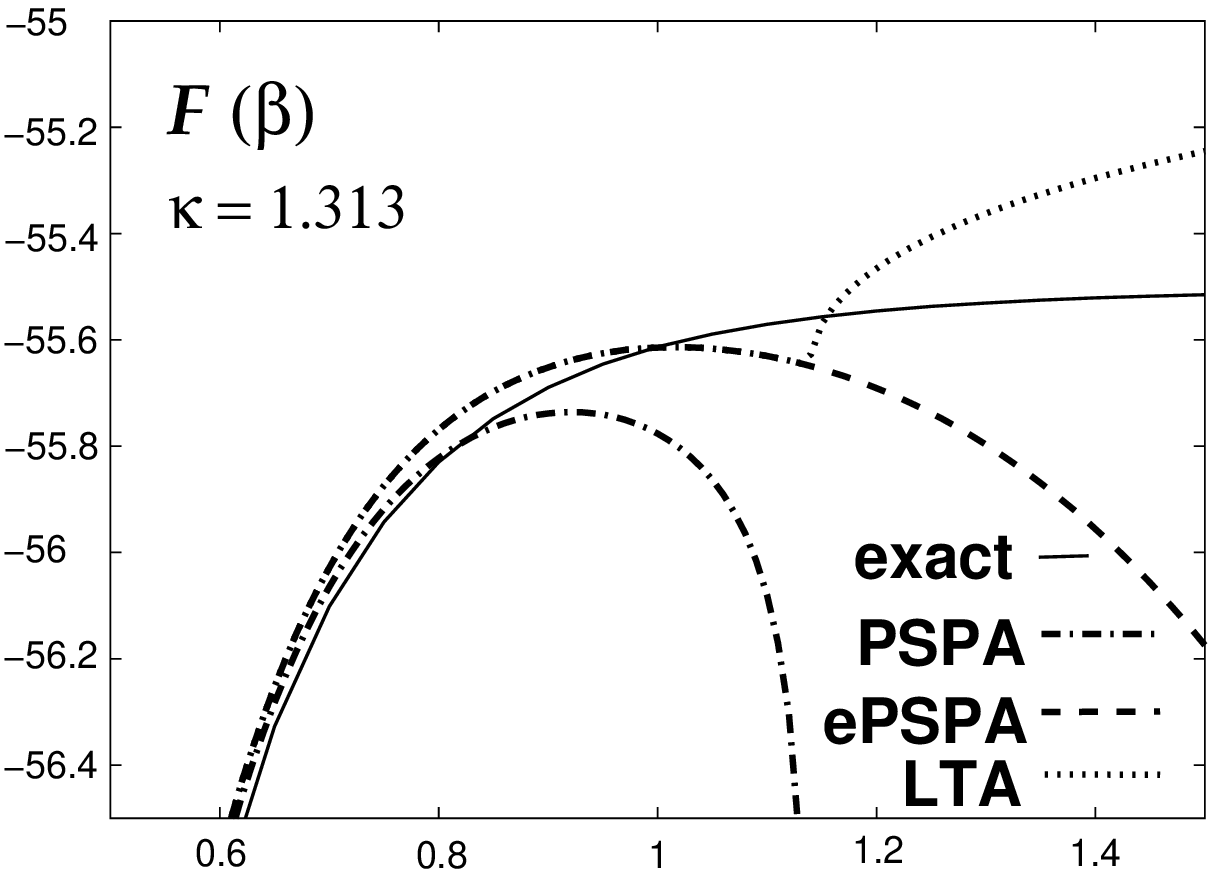}
\includegraphics[width=0.3\linewidth]{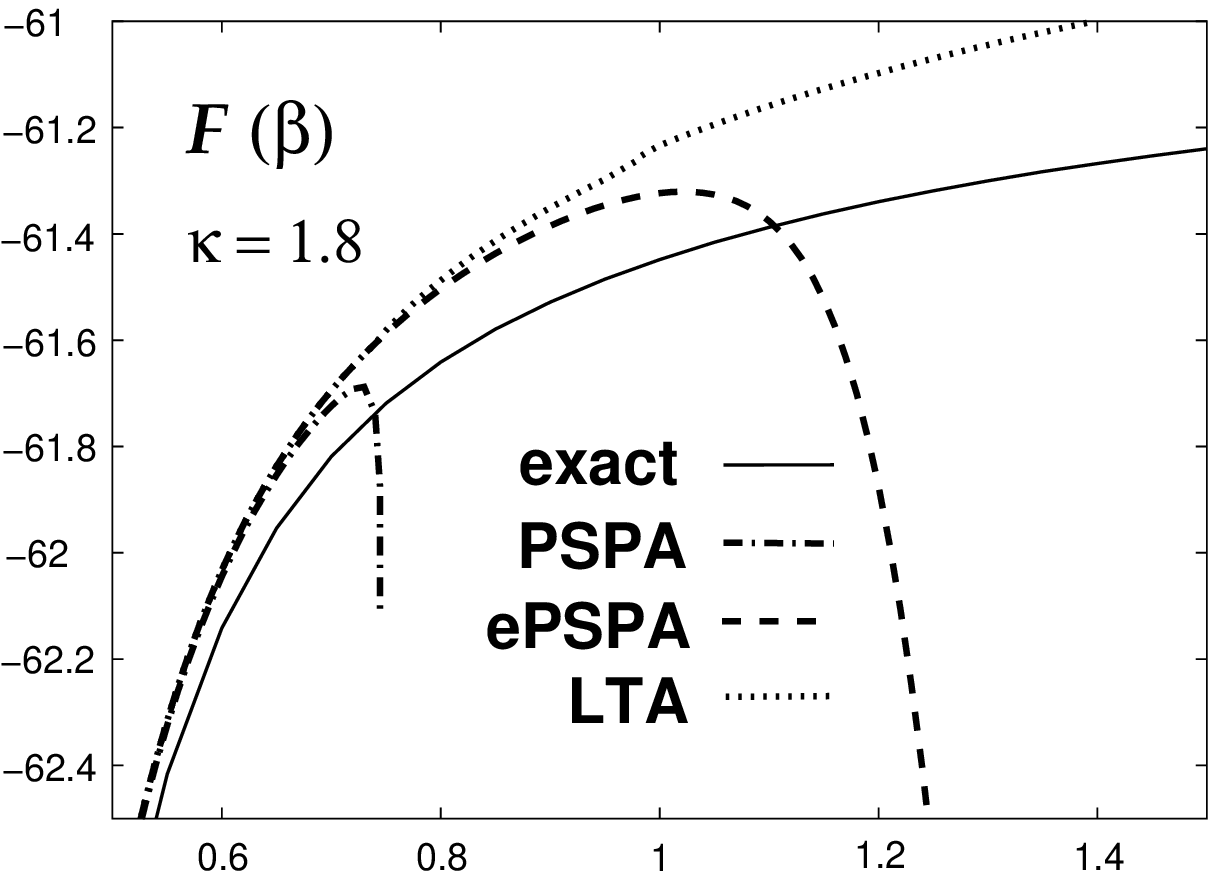}}
\caption{\label{fig-Fbeta} 
Blow up of the crossover region for ${\cal F}(\beta)$ 
of the Lipkin-Meshkov-Glick model (\ref{lip_basHam}) 
within various approximations. Parameters are as in
(\ref{paramPBB}). For $\kappa = 0.8$ no phase transition occurs: 
$T_{0} = 0$. For $\kappa = 1.313$ the inverse crossover temperature is 
$\beta_{0} = 1.1392 \,\textrm{MeV}^{-1}$ whereas for $\kappa = 1.8$ 
the corresponding value is $\beta_{0} = 0.7433 \,\textrm{MeV}^{-1}$.}
\end{figure*}

In figs.~\ref{fig-Fbeta1313enl}--\ref{fig-Fbeta}c we present plots of the 
free energy (\ref{FfromZ}) as a function of the inverse temperature $\beta$. 
Fig.~\ref{fig-Fbeta1313enl} shows 
the global behavior for $\kappa = 1.313$ (the value used in 
\cite{pug.bop.brr:ap:91}).
Figs.~\ref{fig-Fbeta}a--c illustrate a blow up of  the region around
$T_0$ for three different coupling strengths. 
The exact result can be obtained by a numerical diagonalization of the
Hamiltonian (\ref{lip_basHam}) in a basis of eigenstates of $\hat{J}_{z}$
or $\hat{J}_{x}$ as explained in 
\cite{pug.bop.brr:ap:91,ath.aly:npa:97,ror.rip:npa:98}.
As expected, pure SPA 
deviates strongly from the exact results with decreasing temperature. 
For $\kappa < 1$ no phase transition occurs and the difference between 
conventional and extended PSPA is very small [see fig.~\ref{fig-Fbeta}a]. 
The inclusion of quantum effects on the RPA level in the conventional PSPA 
(\ref{defCPSPA}) improves the results a lot as long as $\beta$ is not too 
close to the breakdown value $\beta_{0} = 1 / T_{0}$ 
[see fig.~\ref{fig-Fbeta1313enl}]. Here the conventional 
PSPA fails as the correction factor diverges for $\kappa > 1$. 
Instead, the extension (\ref{C-extPSPA}) smoothly passes the crossover region
and agrees very well with exact results even for inverse temperatures
slightly above $\beta_{0}$.  The ePSPA definitely breaks
down near $\beta = 2\beta_{0}$ where $\lambda_2=0$. The LTA 
[see Sec.~\ref{lowtemp}] smoothly matches with the ePSPA somewhat below $
\beta_{0}$ and is able to 
cover even the range $\beta \gg \beta_{0}$ with astonishing accuracy.

One can also see in fig.~\ref{fig-Fbeta}b that the matching
between ePSPA and LTA even though it appears
to be continuous in temperature, may lead to discontinuities in first and
higher order derivatives. The latter ones produce unphysical
results for the internal energy and specific heat (see below) around the 
matching point close to $T_0$. 
This kind of behavior is typical for the matching
between different semiclassical approximations each one designed for different
ranges in parameter space. It can be overcome in principle by invoking
 uniform semiclassical approximations. Apart from that the ePSPA together 
with the LTA provide a practicable and very efficient way to evaluate 
the free energy over a broad temperature range.

\subsection{Internal energy and specific heat}
\label{inten}

\begin{figure}
\resizebox{0.45\textwidth}{!}{\includegraphics{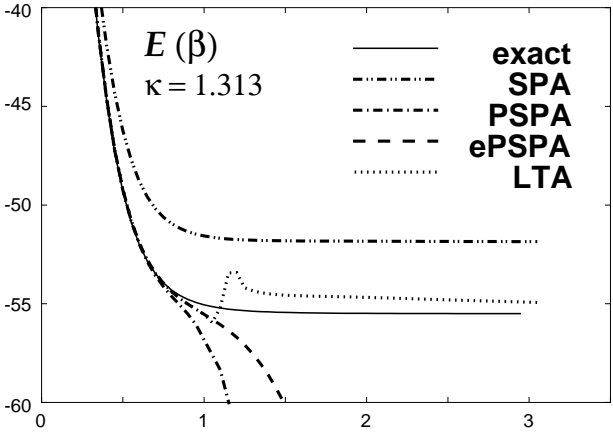}}
\caption{\label{fig-Ebeta} 
${\cal E}(\beta)$ of the Lipkin-Meshkov-Glick model (\ref{lip_basHam}) 
for the parameters of (\ref{paramPBB}) and 
$\kappa = 1.313$. The phase transition occurs at 
$\beta_{c} = 0.4 \,\textrm{MeV}^{-1}$ and the inverse crossover 
temperature is $\beta_{0} = 1.1392 \,\textrm{MeV}^{-1}$.}
\end{figure}

The internal energy 
(comp. \cite{pug.bop.brr:ap:91})
\bel{EfromZ}
{\cal E}(\beta) = 
-\frac{\partial}{\partial \beta} \ \textrm{ln} \,\mathcal{Z}(\beta)
\end{equation}
is shown in fig.~\ref{fig-Ebeta}. 
In the crossover region $\beta \approx \beta_{0}$ the extended PSPA 
again is an improvement over conventional PSPA but tends to fail
for larger $\beta$. The low temperature approximation of Sec.~\ref{lowtemp} 
delivers reasonable values for $\beta > \beta_{0}$. As already 
mentioned above, it shows unphysical discontinuities close to 
$\beta_{0} = 1/T_{0}$.

\begin{figure}
\resizebox{0.45\textwidth}{!}{\includegraphics{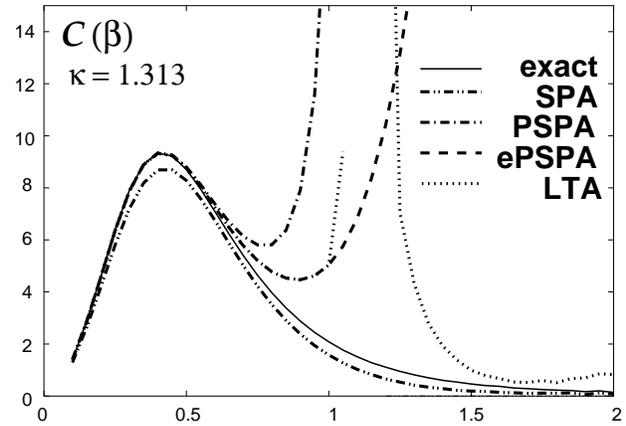}}
\caption{\label{fig-Cbeta} 
${\cal C}(\beta)$ of the Lipkin-Meshkov-Glick model (\ref{lip_basHam}) 
for the parameters of (\ref{paramPBB}) and 
$\kappa = 1.313$. The phase transition occurs at 
$\beta_{c} = 0.4 \,\textrm{MeV}^{-1}$ and the inverse crossover 
temperature is $\beta_{0} = 1.1392 \,\textrm{MeV}^{-1}$.}
\end{figure}

The specific heat of the system (comp. \cite{pug.bop.brr:ap:91}) 
\bel{CfromZ}
\mathcal{C}(\beta) = \beta^{2} \ \frac{\partial^{2}}{\partial \beta^{2}} 
\ \textrm{ln} \,\mathcal{Z}(\beta)
\end{equation}
is seen in fig.~\ref{fig-Cbeta}.  Pure SPA turns out to provide a 
reasonable {\em global} approximation of the specific heat over a broad 
range of temperatures, whereas the conventional and extended PSPA lead 
to very good results at higher temperatures,  smaller $\beta$,  
but deviate already for $\beta < \beta_{0}$. 
The LTA of Sec.~\ref{lowtemp} again shows a discontinuity at 
$\beta = \beta_{0}$ and supplies a reasonable estimate for 
$\beta > \beta_{0}$. 

\subsection{Level densities}
\label{levden}

\begin{figure}
\resizebox{0.45\textwidth}{!}{\includegraphics{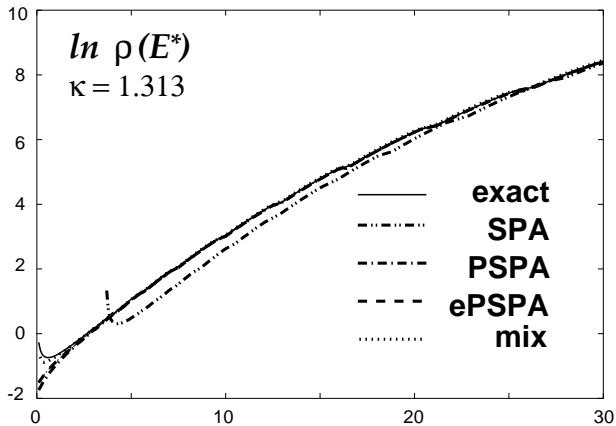}}
\caption{\label{fig-rhoE} 
$\textrm{ln} \,\rho ({\cal E}^{*})$ of the Lipkin-Meshkov-Glick model 
(\ref{lip_basHam}) 
for the parameters of (\ref{paramPBB}) and $\kappa = 1.313$. ${\cal E}^{*}$ 
is measured in MeV.}
\end{figure}

\begin{figure}
\resizebox{0.45\textwidth}{!}{\includegraphics{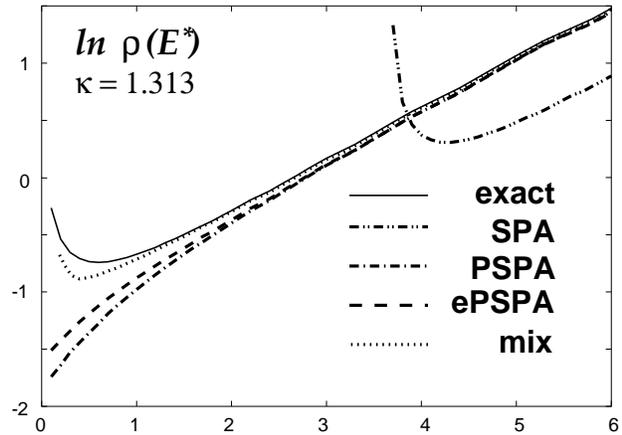}}
\caption{\label{fig-rhoE-low} 
Magnification of fig.~\ref{fig-rhoE} for small excitations.}
\end{figure}

As already mentioned in the introduction the development of the conventional 
PSPA was motivated to some extent by the need to calculate nuclear
level densities as a function of the excitation energy ${\cal E}^{*}$. 
Given the partition function this can be achieved by an inverse Laplace 
transform. Within a saddle point approximation (Darwin-Fowler method) 
one finds the inverse thermal temperature $\beta^{*}$ from
${\cal E}^{*}={\cal E}(\beta^{*}) - {\cal E}_{0}$ and arrives at the 
following formula for the level density \cite{bohra.mottelsonb.1}:
\bel{rhofromZ}
\rho({\cal E}^{*}) = \frac{1}{\sqrt{2\pi \,{\cal D}}} \ \exp
[\mathcal{S}(\beta^{*}) - \beta^{*} \mathcal{E}_{0}]\, .
\end{equation}
The entropy is given by $\mathcal{S}(\beta) = \beta \,[{\cal E}(\beta) - 
{\cal F}(\beta)]$ and
\bel{defD}
{\cal D} = 
\left| \left( \frac{\partial^{2}}{\partial \beta^{2}} \ \textrm{ln} 
\,\mathcal{Z}(\beta) \right)_{\beta = \beta^{*}} \right| = 
\left| \frac{1}{(\beta^{*})^{2}} \,\mathcal{C}(\beta^{*}) \right| \ .
\end{equation}
Results are depicted in fig.~\ref{fig-rhoE}.
Here, high excitation energies correspond to high temperatures. For
this reason approximations of SPA type are expected to be exact at large 
${\cal E}^{*}$ as confirmed in fig.~\ref{fig-rhoE}. For low excitations the
situation is different.  
The pure SPA is not able to deliver results for ${\cal E}^{*} 
\lesssim 4 \,\textrm{MeV}$. The reason is simply that the SPA result for the 
internal energy is about $4 \,\textrm{MeV}$ larger than the exact result at 
small temperatures [see fig.~\ref{fig-Ebeta}] and therefore the condition 
${\cal E}(\beta^{*}) - {\cal E}_{0} = {\cal E}^{*} \lesssim 4 \,\textrm{MeV}$ 
cannot be fulfilled within SPA. The conventional and extended PSPA 
results agree with the exact ones much better than those of pure SPA. 
Essential deviations only occur for ${\cal E}^{*} \lesssim 2 \,\textrm{MeV}$ 
and the extended version is a little better than the conventional one 
[see fig.~\ref{fig-rhoE-low}]. 

The main problem of the conventional and extended PSPA with respect to the 
level density can be traced back to the quantity ${\cal D}$ of (\ref{defD}) 
that is proportional to the specific heat $\mathcal{C}(\beta^{*})$ and 
enters in the denominator of (\ref{rhofromZ}). Divergencies in the 
specific heat as seen for PSPA and ePSPA in fig.~\ref{fig-Cbeta}
therefore have a tremendous impact on the results for 
the level  density. Therefore both types of PSPA deliver level
densities that are too small which becomes significant at small 
excitations corresponding to $\beta \gtrsim \beta_{0}$.  
The following strategy seems to be reasonable: In order to calculate the 
level density using (\ref{rhofromZ}) we exploit an approximation for the 
specific heat that is {\em globally} reasonable and combine it with the best 
approximation at hand for the internal and free energy. That means to
use {\em pure SPA} (see fig.~\ref{fig-Cbeta}) for the specific heat and 
the ePSPA/LTA for the remaining quantities. 
As shown in fig.~\ref{fig-rhoE-low} this procedure improves the conventional 
and extended PSPA a great deal at low excitations. Remarkably, it even
describes the bending up of the level density qualitatively correct.

\section{Conclusion}
\label{concl}

We have applied the path integral approach to
approximately evaluate the partition function of a finite interacting
many body system in the low temperature regime. This requires to
extend the conventional PSPA down to temperatures below the crossover
temperature $T_0$. At these temperatures the simple mean field solution
becomes unstable and large quantum fluctuations arise. 
 The crucial step is to go beyond the Gaussian approximation on which
the PSPA relies and take into account 
anharmonicities of certain critical vibrational modes. This
procedure stabilizes the semiclassical type of approximation to the
path integral of the partition function and holds true down to temperatures 
 $T \approx T_{0}/2$. There, an instability in another direction in
functional space shows up, which successively proceeds to occur at all
$T_{r} = T_{0}/r$ ($r>1$). In order to treat the temperature range $T\ll T_0$,
but still sufficiently above $T=0$, we proposed an approximation where
only mean fields with small quantum fluctuations around the
minima of the static free energy are taken to contribute.

In this way, we could study thermodynamic properties of the archetypical
Lipkin-Meshkov-Glick model even far below the crossover temperature. 
In particular,
for the free and internal energies results were gained, that agree 
very well to the exact ones. Combining our extensions to the PSPA with 
pure SPA we obtained  very accurate  results for the level density even at
low excitation  energies. 

Improvements beyond the PSPA  for $T \approx T_{0}$ and $T \ll T_{0}$ 
are desirable in order to get reasonable approximate results for many
body systems 
in the low temperature range where usual  Monte Carlo techniques
become very expensive. This is particularly true as often the
critical temperature $T_{c}$  
where a  phase transition takes place and the crossover temperature  
$T_{0}$ where the PSPA breaks down are of the same order of
magnitude. For instance, for finite nuclei one finds for the nucleus
$^{164}\textrm{Er}$ that $T_{c} \approx 0.45 \,\textrm{MeV}$, 
while PSPA is reliable only for $T > 0.25 \,\textrm{MeV}$ 
\cite{ror.can.rip:ap:99}. 
In \cite{agb.ana:plb:98} the range of applicability of PSPA is given by 
$T > 0.2 \,\textrm{MeV}$ for the nuclei $^{104}\textrm{Pd}$ and 
$^{114}\textrm{Sn}$. 

In general,  for these or even lower temperatures (of the
order of  $10^{-1} \,\textrm{MeV}$ and below for nuclei)  the 
concept of temperature is at least doubtful for small isolated
systems. This is different for systems on a mesoscopic scale. There 
 the coupling to a macroscopic heat bath is always present such that 
temperature can be fixed from outside. Further, the number of  
constituents is typically much larger as e.g.\ for finite nuclei.
Accordingly, as addressed above, the conventional PSPA has 
 been applied to small superconducting and superfluid systems. 
For superconducting particles of nanometer scale one roughly finds 
$T_{0} \approx T_{c}/2$ \cite{rossignoli} so that 
an extension of the PSPA is clearly relevant. 
Work in this direction is in progress.

\begin{acknowledgement}
One of the authors (C.R.) would like to thank H.~Hofmann for critical 
 and fruitful discussions and A. Ansari for helpful comments. 
C.R. and J.A. thank for the kind hospitality at the $\textrm{ECT}^{*}$, 
Trento, Italy, during  the workshop and collaboration meeting 
``Transport in Finite Fermi Systems''  in May 2000 where part of this
work was done.  
\end{acknowledgement}

\end{document}